\documentclass[twocolumn,showpacs,preprintnumbers,amsmath,amssymb,showkeys]{revtex4}

\usepackage{graphicx}% Include figure files
\usepackage{dcolumn}% Align table columns on decimal point
\usepackage{bm}% bold math
\usepackage{bezier}

\begin{document}

\title{Andreev levels as a quantum dissipative environment}
\author{Artem V. Galaktionov$^{1}$, Dmitry S. Golubev$^{2}$
and Andrei D. Zaikin$^{3,1}$
}
\affiliation{$^1$I.E.Tamm Department of Theoretical Physics, P.N.Lebedev
Physical Institute, 119991 Moscow, Russia\\
$^2$Low Temperature Laboratory, Department of Applied Physics, Aalto University, Espoo, Finland\\
$^3$Institute of Nanotechnology, Karlsruhe Institute of Technology (KIT), 76021 Karlsruhe, Germany
}

\begin{abstract}
We argue that at subgap energies quantum behavior of superconducting weak links can be exactly
accounted for by an effective Hamiltonian for a Josephson particle in a quantum dissipative environment formed by Andreev levels. This environment can constitute an important source for intrinsic inelastic relaxation and dephasing in highly transparent weak links. We investigate the problem of macroscopic quantum tunneling in such weak links demonstrating that -- depending on the barrier transmission -- the supercurrent decay can be described by three different regimes: ($i$) weak intrinsic dissipation, ($ii$) strong intrinsic dissipation and ($iii$) strong capacitance renormalization. Crossover between quantum and thermally-assisted supercurrent decay regimes can also be strongly affected by the Andreev level environment.

\end{abstract}


\maketitle

\section{Introduction}

Quantum dissipation is known to occur as a result of interaction with an effective environment. Quite generally,
this environment can be modeled as a set of harmonic oscillators. Tracing out the oscillator degrees of freedom one
naturally arrives at the Feynman-Vernon influence functional theory \cite{FH} and the Caldeira-Leggett description of
quantum dissipation \cite{CL,Weiss}. Electrons in metallic conductors can also play the role of a quantum dissipative environment,
as it is illustrated, e.g., by the Ambegaokar-Eckern-Sch\"on (AES) effective action treatment of metallic tunnel junctions
\cite{AES,SZ}. Further extension of the influence functional technique also allows to directly account for Fermi statistics
\cite{GZ99} in the situation when electrons in a metal form an effective environment "for themselves".

In superconducting tunnel junctions dissipation at low enough temperatures/energies can only occur {\it extrinsically}
(e.g. by attaching an Ohmic shunt resistor \cite{CL,Weiss,SZ}) as there exist no states with energies
below the superconducting gap $\Delta$ in such junctions.
The situation changes if one goes beyond the tunneling limit. In this case
subgap Andreev bound states \cite{And} are formed inside superconducting weak links. For sufficiently short junctions
the corresponding bound state energies are $\pm\epsilon_n(\varphi )$, where
\begin{equation}
\epsilon_n(\varphi)=\Delta\sqrt{1-T_n\sin^2(\varphi/2)},
\label{And}
\end{equation}
$T_n$ denote normal transmissions of conducting channels and $\varphi$ is the superconducting phase difference
across the junction. While in the tunneling limit $T_n \ll 1$
one has $\epsilon_n(\varphi)\simeq \Delta$ for any value $\varphi$, at higher
transmissions the energies of Andreev levels (\ref{And}) can drop well below
$\Delta$ and may even tend to zero for fully open channels and $\varphi \approx \pi$.

Here we will demonstrate that Andreev bound states -- along with model oscillators \cite{FH,CL,Weiss} or electrons in a metal \cite{AES,SZ,GZ99} --
may act as {\it intrinsic} quantum dissipative environment for the Josephson phase $\varphi$ strongly affecting quantum properties of
superconducting weak links.

\section{Effective Hamiltonian for a weak link} 

In what follows we will consider a current biased superconducting junction characterized by a geometric capacitance $C_0$ and an arbitrary distribution of normal transmissions $T_n$ among ${\cal N}$ transport channels. Provided the phase $\varphi$ does not fluctuate the junction may conduct the supercurrent \cite{KO}
\begin{eqnarray}
I_{S}(\varphi )=\frac{e\Delta^2\sin\varphi}{2}\sum_n\frac{T_n}{\epsilon_n(\varphi)}\tanh\frac{\epsilon_n(\varphi)}{2T}.
\label{Ichi}
\end{eqnarray}
Here and below the sum runs over all channels from $n=1$ to $n={\cal N}$. In order to describe quantum fluctuation effects it is necessary to treat the phase $\varphi$ as a quantum variable \cite{SZ}. The generalization of the AES type-of-approach can be worked out also beyond the tunneling limit employing
both Matsubara \cite{Z} and Keldysh \cite{SN} techniques, however the resulting effective action becomes tractable only in certain physical situations.

One of such situations is realized if phase fluctuations remain sufficiently weak \cite{we,we2}. Splitting the phase variable into constant and
fluctuating parts $ \varphi (t)=\chi +\phi (t)$ and assuming $|\phi (t)| \ll 1$ one can express the kernel of the Keldysh evolution operator $J$
as a double path integral
\begin{equation}
J=\int {\cal D}\phi_F{\cal D}\phi_B \exp (iS_0[\phi_F]-iS_0[\phi_B]+iS_{\rm R}-S_{\rm I}),
\end{equation}
where the phase variables $\phi_{F,B}$ are defined respectively on the forward and backward branches of the Keldysh contour and $S_0[\phi_{F,B}$] define
local in time contributions
\begin{equation}
S_0 [\phi ]=  \int\limits_0^t dt'\left[\frac{C\dot\phi ^2}{8e^2}-U(\chi+\phi(t'))\right].
\label{S0Jp}
\end{equation}
Here $C$ is the effective junction capacitance which may differ from $C_0$ due to retardation effects \cite{we} and
\begin{equation}
U(\varphi )=-2T\sum_{n}\ln \left[\cosh \frac{\epsilon_n(\varphi )}{2T}\right]-\frac{I\varphi }{2e}
\label{Uchi}
\end{equation}
is the effective potential where the first term is recovered by integrating the supercurrent (\ref{Ichi}) over $\varphi$ and the
second term accounts for the bias current $I$.

The remaining -- nonlocal in time -- terms
\begin{eqnarray}
S_{\rm R}&=&\int\limits_0^t dt'\int\limits_{0}^{t}dt''
{\cal R}(t'-t'') \phi_-(t')\phi_+(t''), \label{SRRR}\\
S_{\rm I}&=&\int\limits_{0}^{t}dt'\int\limits_{0}^{t}dt''
{\cal I}(t'-t'') \phi_-(t')\phi_-(t'')
\label{SIII}
\end{eqnarray}
describe the influence functional for the phase variable. Here we denote $\phi_+=(\phi_F+\phi_B)/2$ and $\phi_-=\phi_F-\phi_B$. Both kernels ${\cal R}(t)$ and ${\cal I}(t)$ in Eqs. (\ref{SRRR}), (\ref{SIII}) are real functions related to each other via the fluctuation-dissipation theorem. General expressions for each of these kernels remain rather involved \cite{we} and contain three
different contributions originating from (i) the subgap Andreev bound states, (ii) the quasiparticle states above the gap and (iii) the interference between (i) and (ii).

Significant simplifications occur provided both temperature and typical phase fluctuation frequencies remain well in the subgap range, i.e. $T, \omega_{\phi} \ll \Delta$.
Under this condition the contributions (ii) and (iii) to the kernel ${\cal I}$ vanish, while the analogous
 contributions to the Fourier component of ${\cal R}$ can be expanded in powers of $\omega /\Delta$ up to terms $\sim \omega^2$
 which yields an effective capacitance renormalization. In the interesting for us limit $1-T_n \ll 1$ and $\pi -\chi \ll \pi$
this renormalization yields
\begin{equation}
C\simeq C_0+e^2{\cal N}/(4\Delta).
\label{capren0}
\end{equation}
What remains is to account for the Andreev terms (i) in both kernels
${\cal R}$ and ${\cal I}$. Making use of the results \cite{we,we2}, we obtain
\begin{eqnarray}
{\cal R}(t)=\sum_n\frac{\gamma_n}{2\epsilon_n(\chi )}\,\theta(t) \sin(2\epsilon_n(\chi ) t),
\label{R111}\\
{\cal I}(t)=\sum_n\frac{\gamma_n}{8\epsilon_n(\chi )}\coth\left(\frac{\epsilon_n(\chi )}{T}\right) \cos \left( 2\epsilon_n(\chi ) t\right),
\label{I111}
\end{eqnarray}
where
\begin{equation}
\gamma_n=T_n^2(1-T_n) \frac{\Delta^4}{\epsilon_n(\chi)}\sin^4\frac{\chi}{2}\tanh\frac{\epsilon_n(\chi)}{2T}.
\label{randg}
\end{equation}
Note that in the limit $\omega \ll \Delta$ and provided $\epsilon_n(\chi ) \sim \Delta$ the Fourier component of the kernel ${\cal R}(t)$ in Eq. (\ref{R111}) can also be expanded in powers of $\omega /\epsilon_n$ up to terms $\sim \omega^2$ giving rise to extra renormalization of $C$ \cite{we} (see below).
For $\epsilon_n(\chi ) \ll \Delta$, however, this expansion is not justified anymore. For this reason in what follows we will keep the kernel ${\cal R}(t)$
in its exact form (\ref{R111}).

It is interesting to observe that the influence functional defined in Eqs. (\ref{SRRR})-(\ref{I111})
is exactly equivalent to that produced by a bath of harmonic oscillators \cite{FH,CL,Weiss} coupled linearly to the fluctuating part of the phase $\phi$. In other words, our weak link can also be described by an effective low energy Hamiltonian
\begin{eqnarray}
\hat H&=& -\frac{2e^2}{C}\frac{\partial^2}{\partial \phi^2}+U(\chi+\phi)\nonumber\\
&&+\sum_n \left[\frac{\hat
P_n^2}{2M_n}+\frac{M_n\omega_n^2}{2}\left(Q_n-\frac{c_n\phi}{M_n\omega_n^2}\right)^2\right].
\label{HQintphi}
\end{eqnarray}
The first and the second lines of  Eq. (\ref{HQintphi}) account respectively for the "Josephson particle"  and for its Andreev level
environment which consists of ${\cal N}$ harmonic oscillators with frequencies $\omega_n=2\epsilon_n(\chi )$
coupled to the "particle coordinate" $\phi$. The coupling constants $c_n$ are identified by the condition $c_n^2/M_n=\gamma_n$.
It is important to emphasize that the Hamiltonian (\ref{HQintphi}) follows directly from a fully microscopic effective action
\cite{we} without any model assumptions.

Before dwelling into further calculations it is instructive to also construct the grand partition function for our weak link
${\cal Z}={\rm Sp} e^{-\hat H/T}$. Making use of the effective Hamiltonian in Eq. (\ref{HQintphi}) we can express ${\cal Z}$ in terms of the path integral
over both $\phi$ and the oscillator coordinates $Q_n$. Integrating out
all $Q_n$-variables we obtain
\begin{equation}
{\cal Z}=\int {\rm D}\phi
\exp \left(- S_{\rm eff}[\phi (\tau)]\right),
\label{Z177}
\end{equation}
where
\begin{eqnarray}
S_{\rm eff}=\int_0^{1/T} d\tau \left[\frac{C\dot\phi ^2}{8e^2}+U(\chi+\phi(\tau))\right]\nonumber\\
+\int\limits_0^{1/T} d\tau_1 \int\limits_0^{1/T} d\tau_2 Y(\tau_1-\tau_2)\phi(\tau_1)\phi(\tau_2)
\label{Seff17}
\end{eqnarray}
is the imaginary time effective action for our superconducting contact and
\begin{equation}
Y(\tau)=\sum_n\frac{\gamma_n}{8\epsilon_n}\left( \frac{\delta(\tau)}{\epsilon_n}-\frac{\cosh\left[2\epsilon_n\left( \left|\tau \right|-\frac{1}{2T}\right) \right]}{\sinh[\epsilon_n/T]}\right).
\label{Y17}
\end{equation}
Expanding the kernel (\ref{Y17}) in the Fourier series
$Y(\tau )=T\sum_{\omega_m}Y_{\omega_m}e^{-i\omega_m \tau}$
with $\omega_m =2\pi m T$, we get
\begin{equation}
Y_{\omega_m}(\chi)= \sum_n\frac{\gamma_n(\chi )}{8\epsilon_n^2 (\chi )}\frac{\omega_m^2}{\omega_m^2+4\epsilon_n^2 (\chi )}.
\label{Yo1717}
\end{equation}

\section{Andreev level bath spectrum and inelastic relaxation}

To begin with, we observe that the coupling constants $c_n^2 \propto \gamma_n$ (\ref{randg})
vanish in the limit $T_n\to 1$. Thus, environmental modes corresponding to fully open transport channels are totally decoupled from the phase variable $\phi$ and, hence, cannot influence its quantum dynamics.

On the other hand, channels with $T_n < 1$ do affect the behavior of $\phi$. As long as $\omega_\phi$ remains much smaller than $\omega_n$ the $n$-th environmental mode may only contribute to
extra capacitance renormalization. Provided the inequality $\omega_\varphi \ll \omega_n$ is fulfilled for all $n$ we again recover the corresponding result for the renormalized capacitance \cite{we}. In the opposite limit $\omega_\varphi \gg \omega_n$ Andreev levels may already act as a {\it quantum dissipative environment} for the fluctuating phase.  This is because the oscillators can get excited to higher energy states as a result of their interaction with $\varphi (t)$. Note that since phase fluctuations remain small the conjugate charge variable, in contrast, fluctuates strongly implying that multiple electron charge transfer is possible through each conducting channel. Accordingly, many of such electrons can in general get excited to the higher of the
two Andreev levels while passing through the $n$-th channel. These processes translate into the excitation of harmonic oscillators to higher levels illustrating the physical reason why the effect of Andreev doublets is equivalent to that of such oscillators.

The frequency spectrum of this quantum dissipative environment depends on the particular distribution of normal transmissions $T_n$. Let us introduce
\begin{eqnarray}
J(\omega,\chi) = \frac{\pi}{2}\sum_n\frac{\gamma_n}{\omega_n}\delta(\omega-\omega_n), \quad \omega_n=2\epsilon_n(\chi ).
\end{eqnarray}
For a junction with large number of channels ${\cal N} \gg 1$ and with transmissions distributed in the interval $0<T_n<1$ with the probability $P(T_n)$ 
after a simple algebra for $\omega_{\min}(\chi)< \omega <2\Delta$ we get
\begin{eqnarray}
J(\omega,\chi)= \frac{\pi (\Delta^2-\omega^2/4)}{2\omega} P(T_\omega)T_\omega(1-T_\omega )\tanh\frac{\omega}{4T}
\label{J_om}
\end{eqnarray}
and $J(\omega,\chi)=0$ otherwise. Here we define
\begin{equation}
T_\omega=\frac{1-\omega^2/(2\Delta )^2}{\sin^2(\chi/2)}, \quad \omega_{\min}(\chi)=2\Delta|\cos (\chi/2)|.
\label{ominmax}
\end{equation}

Intrinsic dissipation due to subgap Andreev levels can be identified if we consider, e.g., the
two lowest energy levels in the Josephson potential well $U(\varphi )$ which can also be treated as a qubit.
The corresponding inelastic relaxation rate for such a qubit $\gamma_{\rm in}(\chi)$ reads
\begin{eqnarray}
\gamma_{\rm in}=\frac{E_C }{\omega_0}J\big(\omega_0,\chi\big)\coth\frac{\omega_0}{2T}.
\label{gammain}
\end{eqnarray}
Here $E_C=e^2/2C$ is the junction charging energy, $\omega_0=\sqrt{8E_CU''(\chi )}$ is the plasma oscillation frequency and the phase $\chi$ is fixed by the condition $I_S (\chi )=I$.

In many cases $\omega_{\min}$ remains of order $\Delta$. E.g., in diffusive junctions at $T \to 0$ and $I$ close to $I_C$ one
finds $\omega_{\min}\simeq 1.1 \Delta$. Only provided $\chi$ is close to $\pi$ and, in addition, there exist transmitting
channels with $T_n \approx 1$ the frequency  $\omega_{\min}$ can go well below $\Delta$. This situation occurs, e.g., in the problem of
macroscopic quantum tunneling (MQT) to be addressed below.

\section{MQT in highly transparent weak links} 

Provided a superconducting weak link with ${\cal N} \gg 1$ is biased by the current $I$ close to $I_C$
the zero resistance state becomes unstable and can decay into a resistive state due to quantum tunneling of the phase across the potential barrier
$U(\varphi )$. In the case of superconducting tunnel barriers with $T_n \ll 1$ this MQT problem with extrinsic dissipation
was thoroughly studied in the past \cite{CL,Weiss}. Here we consider the opposite limit of highly transparent weak links with $T_n \approx 1$ in which case
the critical current $I_C=I_S(\chi_c )$ is reached at the phase value $\chi=\chi_c$ close to $\pi$. 
Such weak links can now be fabricated in a controlled manner employing a variety of different materials including, e.g., atomic point contacts \cite{exp1}, 
graphene-based weak links \cite{exp2,exp3}, high transparency Al/BiTe/Al double barrier heterostructures \cite{exp4}, or InAs nanowire Josephson junctions \cite{exp5}. 
In some of these experiments effective channel transmissions with values very close to unity were demonstrated. 

In what follows for simplicity we will assume that all the junction transmissions have the same value $T_n={\cal T}$
and, hence, $\epsilon_n(\chi )\equiv \epsilon (\chi )$, cf. Eq. (\ref{And}). We also define
the reflection coefficient $r =1-{\cal T} \ll 1$ and the parameter $q(T)=1-I/I_C(T) \ll 1$.

Let us evaluate the supercurrent decay rate with the exponential accuracy $\Gamma \propto \exp (-A)$. At  $T \to 0$ the effective potential (\ref{Uchi})
reduces to a simple form
\begin{equation}
U(\varphi )=-I\varphi /2e-{\cal N}\Delta\sqrt{1-{\cal T}\sin^2(\varphi/2)}.
\label{UT0}
\end{equation}
For $q \ll \sqrt{r}$ Eq. (\ref{UT0}) can be simplified further by expanding it in powers of $\phi$ around
$\chi=\chi_c=\pi - \arccos [(1-\sqrt{r})^2/{\cal T}]$.
Dropping an unimportant constant we get
\begin{equation}
U(\chi_c+\phi )\simeq \frac{\Delta{\cal N}\nu}{2}\left[q\phi -\frac{\phi^3}{6}\right], \quad \nu = 1-\sqrt{r}
\label{phikubs}
\end{equation}
Observing a strong inequality $r \ll 1$ below we will set $\nu \simeq 1$. For
$q \gtrsim \sqrt{r}$ the expansion (\ref{phikubs}) is no more sufficient, and the exact form of $U$ (\ref{UT0})
should be employed.

Provided geometric capacitance $C_0$ is large it suffices to ignore both effects of capacitance renormalization (\ref{capren0}) and of
Andreev levels by formally setting $C=C_0$ and $Y(\tau ) \to 0$ in Eq. (\ref{Seff17}). Then our
MQT problem reduces to that of a quantum particle with mass $C_0/4e^2$ which tunnels under the barrier in the potential (\ref{UT0}).
This problem is resolved easily with the result 
\begin{equation}
A=\frac{36\kappa U_0}{5\omega_0}, \quad \kappa =
\begin{cases}
1, & q \ll \sqrt{r},
\\
\frac{2\sqrt{2}}{3+\sqrt{6}}, & q \gg \sqrt{r},
\end{cases}
\label{expnodiss17}
\end{equation}
where $U_0=\Delta {\cal N}(2q)^{3/2}/3$ defines the potential barrier height and $\omega_0^2 = 2e^2\Delta{\cal N}(2q)^{1/2}/C_0$.
\begin{figure}[h]
\includegraphics[width=0.89\columnwidth]{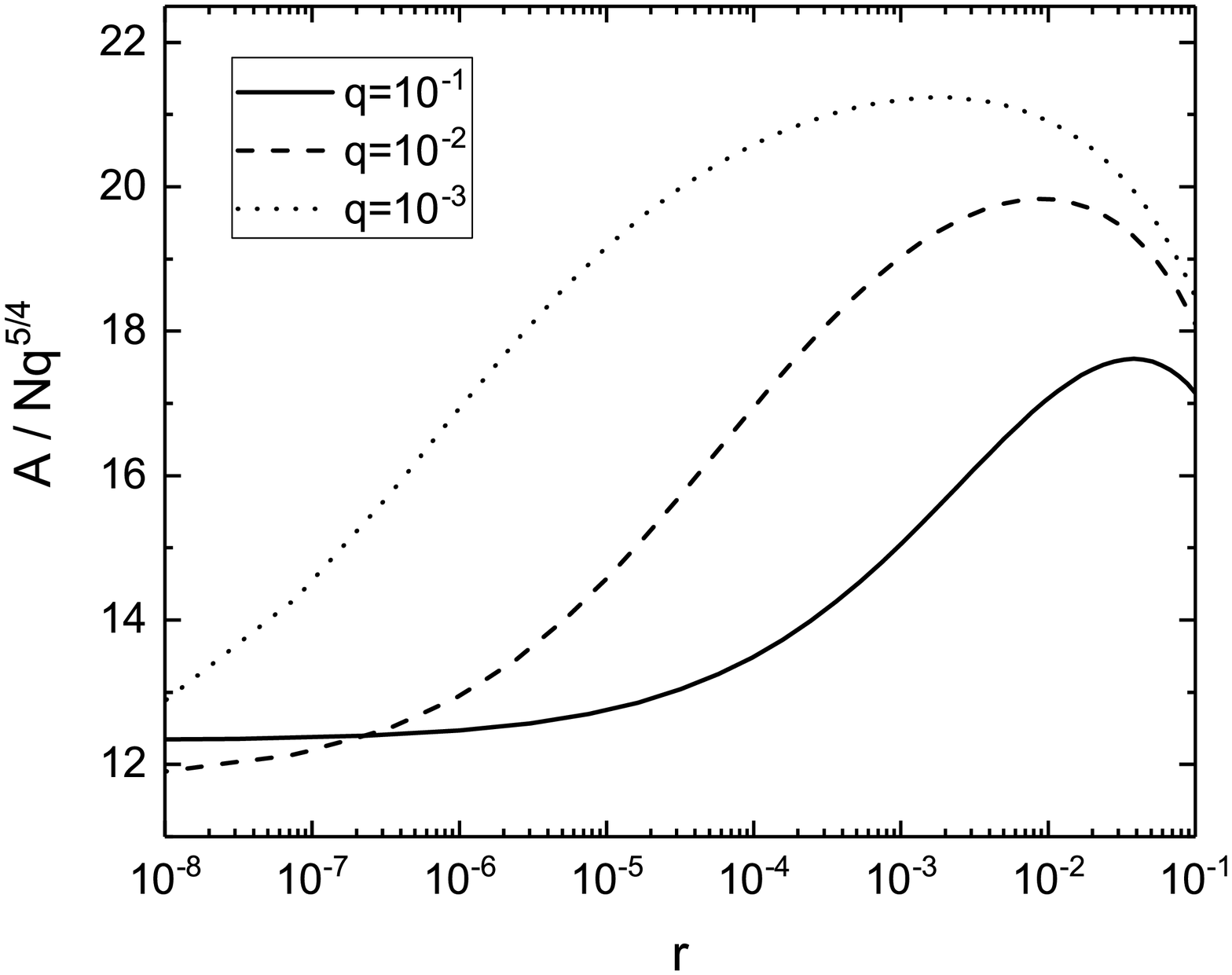}
\caption{The exponential factor $A$ plotted versus $r$ for ${\cal N}=400$ and $\Delta/E_C=2.49\times 10^{4}$, cf. Eq. (\ref{expnodiss17}).}
\label{fig1}
\end{figure}

Note that the numerical prefactor $\kappa$ is almost two times bigger in the limit $q \ll \sqrt{r}$ than in the opposite one $q \gg \sqrt{r}$.
Since $A \gg 1$ we conclude that for $1-{\cal T} \ll 1$ an increase of ${\cal T}$ by a very small value can result in an increase of
the supercurrent decay rate $\Gamma$ by orders of magnitude. This is because the potential barrier $U$ in the limit $q \gg \sqrt{r}$ is
substantially "thinner" than that at $q \ll \sqrt{r}$, while the barrier height $U_0$ remains the same in both limits.
Hence, the tunneling probability can be much bigger in the former limit, see also Fig. 1.

Let us now assume that $C_0$ is small and the capacitance $C$ is dominated by
the last term in Eq. (\ref{capren0}), i.e. we set $C\simeq e^2{\cal N}/(4\Delta )$.
Provided $\sqrt{r} \lesssim q$ the coupling constant (\ref{randg}) between the particle $\phi$ and the Andreev bath is
sufficiently small. Accordingly, dissipation remains weak and can be treated perturbatively. At $\sqrt{r} \ll q$ we find
\begin{equation}
A\simeq\frac{4(3-\sqrt{6})}{5}{\cal N}(2q)^{5/4}\left[1+\alpha \left(\frac{\sqrt{r}}{q}\right)\right],
\label{Aa1777}
\end{equation}
where $\alpha \ll 1$ accounts for dissipation \cite{FNA}. As long as $\sqrt{r}\lesssim q$
the value $A$ grows with $r$ merely due to the potential profile change (as accounted for by
the parameter $\kappa$ in Eq. (\ref{expnodiss17}) and also illustrated in Fig. 1) rather than due to dissipation.
Hence, in the limit $\sqrt{r} \ll q$ the function $\alpha$ can be safely neglected.

For $\sqrt{r}> q$ effective coupling of $\phi$ to the Andreev level bath becomes strong and the last term in Eq. (\ref{Seff17}) should be treated nonperturbatively. It is well known that quantum tunneling can be described in terms of
classical dynamics of the particle $\phi$ propagating in the inverted potential $-U$ along the so-called bounce
trajectory. Identifying the bounce frequency with that of small oscillations $\tilde\omega_{0}$ near the bottom of this potential at $\phi =\sqrt{2q}$ and setting again $C\simeq e^2{\cal N}/(4\Delta )$, after trivial algebra we get
\begin{equation}
\tilde\omega_{0} = \Omega (q,r^{1/4}),
\label{omphi}
\end{equation}
where we introduced the function
\begin{equation}
\Omega (q,y)=2\Delta \sqrt{\frac{Z}{2}+\sqrt{\frac{Z^2}{4}+y^2\sqrt{8q}}}
\label{O0gen}
\end{equation}
with $Z=\sqrt{8q}-y^2-r/y^3$. For the potential (\ref{phikubs}) we find
\begin{equation}
A \simeq 36 U_0/(5\tilde\omega_{0}).
\label{Agen0}
\end{equation}
This formula together with Eqs. (\ref{omphi}) and (\ref{O0gen}) accounts for a trade-off between two different tunneling regimes. Let us define the value $q_c$ from the equation $\Omega (q_c,r^{1/4})=2\epsilon (\chi_c)=2\Delta r^{1/4}$ which yields $q_c=\sqrt{r}/32 + (r^{3/4}+r)/8$. In the adiabatic limit $\omega_\phi \ll 2\Delta r^{1/4}$ (or $q <q_c$) Andreev oscillators are ``fast'' and provide strong capacitance renormalization \cite{FN}
\begin{equation}
C^*=C + e^2{\cal N}/(4\Delta r^{1/4}).
\label{C*new}
\end{equation}
The particle $\phi$ then becomes heavier albeit its energy remains conserved during tunneling and Eq. (\ref{Agen0}) yields
\begin{equation}
A\simeq (6/5)2^{3/4}{\cal N}q^{5/4}r^{-1/8}.
\label{strbath}
\end{equation}
In the opposite antiadiabatic limit $\omega_\phi \gg 2\Delta r^{1/4}$ (or $q >q_c$) Andreev oscillators become ``slow''
generating effective potential renormalization  $U \to U+ \Delta{\cal N}r^{1/4}\phi^2/8$.
Such oscillators can get excited to higher levels taking energy from the tunneling particle $\phi$. Hence, for $q_c <q \lesssim \sqrt{r}$ Eq. (\ref{Agen0}) describes a strong dissipation regime and matches smoothly with Eq. (\ref{Aa1777}) at  $q \sim \sqrt{r}$.

\section{Quantum-to-classical crossover}

Quantum decay can only occur at low enough temperatures $T<T_0$ whereas at $T>T_0$ thermal activation takes over with $\Gamma \propto \exp (-U_0/T)$. In order to analyze the crossover between these two regimes we will employ an approximate
form (\ref{phikubs}) for the potential energy $U$. At lower $T \ll \epsilon (\chi_c)=\Delta r^{1/4}$ Eq. (\ref{phikubs}) holds for $q \lesssim \sqrt{r}$, as we already indicated above. At higher temperatures
$\Delta r^{1/4} \ll T \ll \Delta$ the approximation (\ref{phikubs}) applies without any further restrictions.
Indeed, at such $T$ and for $r \ll 1$ Eq. (\ref{Ichi}) reduces to
\begin{equation}
I_S(\varphi )= e{\cal N}\Delta\sin\frac{\varphi}{2}
\tanh \left(\frac{\Delta}{2T}\cos\frac{\varphi}{2}\right)
\label{IchiHT}
\end{equation}
reaching its maximum $I_S(\varphi )=I_C (T)$ at $\varphi =\chi_c$, where now $\chi_c=\pi -(2T/\Delta )W(2\Delta^2/T^2)$
and $W(z)$ is the Lambert W-function defined by the equation $W\exp (W)=z$.
For $q \ll T^2/\Delta^2$ we can again expand the potential $U$ (\ref{Uchi}) in $\phi=\varphi -\chi_c$ and reproduce Eq. (\ref{phikubs}) with $\nu=1$.

At temperatures in the vicinity of the crossover to thermal activation quantum tunneling is described by the bounce trajectory $\tilde \phi (\tau )$ which remains close to the local maximum of the potential $U$ (\ref{phikubs}) at $\phi=\sqrt{2q}$. The value $T_0$ is formally identified \cite{Weiss,LO83,GW84} as a temperature at which the bounce first reduces to $\tilde \phi (\tau )= \sqrt{2q}$ meaning that $A=U_0/T_0$ at this point. This is achieved provided the corresponding eigenvalue of the operator $\delta^2 S_{\rm eff}/\delta \phi^2_{\phi =\sqrt{2q}}$ vanishes, which yields
\begin{equation}
\pi^2T^2C/e^2+2Y_{2\pi T}(\chi_c)={\cal N}\Delta\sqrt{q/2}.
\label{bouncepoint}
\end{equation}
Resolving Eq. (\ref{bouncepoint}) one determines the crossover temperature $T_0$. If the geometric capacitance $C_0$ is large, it suffices to set $C=C_0$ and $Y_{2\pi T}\to 0$ in Eq. (\ref{bouncepoint}). Then we immediately recover the standard result \cite{Affleck} $T_0=\omega_0 /(2\pi)$, where $\omega_0$ is defined below Eq. (\ref{expnodiss17}). This result holds for $q \lesssim \sqrt{r}$. In the limit  $\sqrt{r} \ll q$ Eq. (\ref{bouncepoint}) cannot be applied anymore.
In this case $U_0/T$ reaches the value $A$ in Eq. (\ref{expnodiss17}) at $T \approx 2T_0$.

\begin{figure}[h]
\includegraphics[width=0.89\columnwidth]{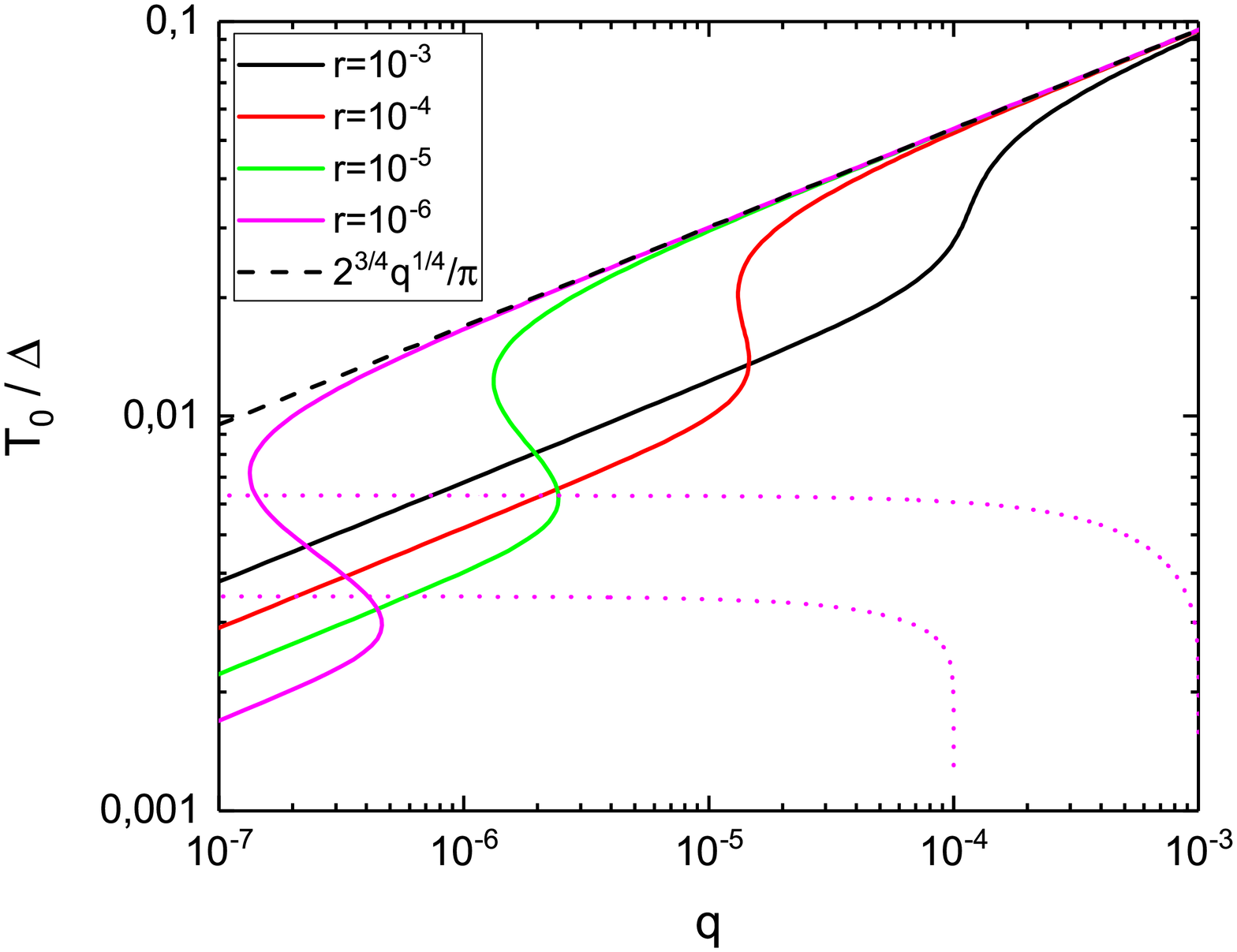}
\caption{(Color online) The function $T_0(q)$ (\ref{T0gen}) for different values of $r$ (solid lines) together with the dependencies $q(T)$ for
$r=10^{-6}$ (dotted lines). The crossover temperature $T_0(I)$ is determined as the intersection point of $T_0(q)$ and $q(T)$.}
\label{fig2}
\end{figure}

Perhaps the most interesting situation occurs if geometric capacitance $C_0$ is small and $C$ is dominated by the last term in Eq. (\ref{capren0}).
Substituting $C=e^2{\cal N}/(4\Delta)$ together with Eq. (\ref{Yo1717}) into Eq. (\ref{bouncepoint}) and resolving the latter with respect to $T$ we obtain
\begin{equation}
T_{0}=\Omega (q,y) /(2\pi), \quad y=\epsilon (\chi_c(T_0))/\Delta .
\label{T0gen}
\end{equation}
In the limit $T_0 \ll \Delta r^{1/4}/\pi$ we have $y\simeq r^{1/4}$ and Eq. (\ref{T0gen}) reduces to $T_0=\tilde\omega_{0}/(2\pi)$. In particular, at small enough $q$ we get $T_{0}=\Delta (8q)^{1/4}r^{1/8}/\pi$.
For $T_0 \gtrsim \Delta r^{1/4}/\pi$ Eq. (\ref{T0gen}) approaches the $r$-independent result $T_{0}=\Delta (8q)^{1/4}/\pi$. The function $T_0(q)$ (\ref{T0gen}) is displayed in Figs. 2 and 3 (inset) for different values of $r$.

Note that for very small $r \lesssim 10^{-4}$ the function $T_0(q)$ becomes multivalued for some values of $q$. This behavior, however, does not imply the presence of several crossover temperatures for a given bias $I$ because the critical current $I_C$ also depends on temperature and, hence, $q=q(T)$. For each value $I$ the crossover temperature $T_0$ should be obtained from the equation
$T_0=T_0(q(T_0))$, as it is also illustrated in Fig. 2. As a result, we arrive at the bias current value $I=I_0(T)$ at which quantum-to-classical
crossover occurs at a given temperature. The function $I_0(T)$ is plotted in Fig. 3 for $r=0.1$ together with $I_C(T)$.
With decreasing $r$ classical activation region shrinks and $I_0(T)$ rapidly approaches $I_C(T)$.

\begin{figure}[h]
\includegraphics[width=0.89\columnwidth]{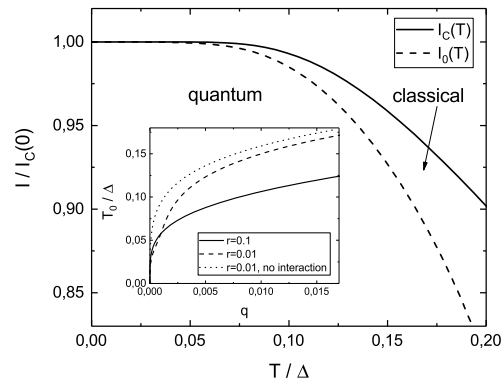}
\caption{The bias current $I_0$ separating the regimes of quantum tunneling and thermal activation
as a function of temperature plotted together with $I_C(T)$ for $r=0.1$. The inset shows the dependence $T_0(q)$ (\ref{T0gen}) for $r=0.1$ and $r=0.01$.}
\label{fig3}
\end{figure}

In summary, we demonstrated that subgap Andreev bound states in superconducting weak links form an intrinsic quantum dissipative environment 
for the Josephson phase $\varphi$ and derived a microscopic low energy Hamiltonian for such weak links. Effective coupling between $\varphi$ and 
Andreev oscillators depends on transport channel transmissions $T_n$ and vanishes for fully open channels with $T_n \to 1$. In the case of highly 
transparent weak links we analyzed both quantum and thermally-assisted decay of the supercurrent and identified the MQT regimes of weak intrinsic 
dissipation, strong intrinsic dissipation and strong capacitance renormalization. Our predictions can be directly tested  
by routinely measuring the statistics of switching currents (see, e.g., recent experiment \cite{Ustinov})
in highly transparent superconducting weak links in combination with Andreev level spectroscopy \cite{Urbina}.


\begin{thebibliography}{20}
\bibitem{FNA} In order to estimate $\alpha$ in Eq. (\ref{Aa1777}) we employ Eq. (\ref{Yo1717}) with $\chi$ corresponding to the local minimum of
the potential (\ref{UT0}). Then we obtain $\alpha (z ) \sim z^2 \ll 1$. Choosing $\chi=\chi_c$ would give $\alpha (z ) \sim \sqrt{z} \ll 1$,
however this choice is inaccurate for $\sqrt{r} \ll q$ and $T \to 0$ since in this limit typical phase fluctuations $\delta \phi \sim \sqrt{q}$ exceed $\pi -\chi_c \sim r^{1/4}$.
\bibitem{FN} The combination of Eqs. (\ref{capren0}) and (\ref{C*new}) is equivalent to the analogous result \cite{we} in the corresponding limit.
\end{thebibliography}
\end{document}